\documentstyle[aps,prl,multicol,epsf]{revtex}
\begin{document}

\title{
Boundary susceptibility in the spin-1/2 chain:\\
Curie like behavior without magnetic impurities
}

\author{Satoshi Fujimoto$^1$ and Sebastian Eggert$^2$} 
\address{
$^1$Department of Physics,
Kyoto University, Kyoto 606-8502, Japan\\
$^2$Institute of Theoretical Physics, Chalmers University of
Technology, S-41296 G\"oteborg, Sweden 
}

\date{\today}
\maketitle
\begin{abstract}
We investigate the low-temperature thermodynamics
of the spin-1/2 Heisenberg chain with open ends.
On the basis of boundary conformal field theory arguments
and numerical density matrix renormalization group calculations, 
it is established
that in the isotropic case the impurity susceptibility 
exhibits a Curie-like divergent behavior as the temperature decreases, 
even in the absence of magnetic impurities.
A similar singular temperature dependence is also found in the boundary
contributions of the specific heat coefficient. 
In the anisotropic case, for $1/2<\Delta<1$, these boundary quantities
still show singular temperature dependence obeying a power law with 
an anomalous dimension.  Experimental consequences will be discussed.
\end{abstract}


%\begin{multicols}{2}

Impurity doping in low dimensional antiferromagnetic systems has been
a topic of great interest in recent years\cite{batis,sach,ess,as,fra,fuji2,egg,qin,bru,rom,haas}. 
New interesting boundary 
phenomena give a better physical insight to the underlying structure of the 
strongly correlated states in those systems.  In 
one-dimensional  systems impurities typically cut the chains and play
the role of effective boundary conditions.  The open ends 
give rise to intriguing boundary phenomena, which result in a
characteristic temperature dependence of the excess susceptibility and
specific heat due to the impurities.  Famous examples include the spin-1
chain where effective spin-1/2 degrees of freedom are created near the 
open ends which show clear experimental 
signatures\cite{batis}.  Even
in the two dimensional Heisenberg model it has been postulated that 
non-magnetic impurities
may give rise to a divergent Curie-like impurity susceptibility\cite{sach}. 
However, the temperature dependence of the boundary susceptibility 
has been much less clear in the case of the antiferromagnetic spin-1/2 chain, 
which is possibly the oldest and most studied prototype of a strongly 
correlated system.  According to the Bethe ansatz solutions for
integrable systems with open boundaries,
the boundary part of the uniform spin susceptibility at {\it zero} temperature
behaves like $\sim 1/[h\{\ln(h)\}^2]$ for a small magnetic field $h$
in the case of SU(2) spin rotational symmetry\cite{ess,as,fra,fuji2}. 
This divergent behavior implies that
in the vicinity of the open edges spin excitations are very sensitive to
small external fields.  It seems therefore likely that the susceptibility 
should also show divergent behavior for small temperatures at zero field.
On the other hand, irrelevant boundary operators are known to only 
produce a small finite impurity susceptibility at low temperatures\cite{egg}.  
We now show that an addition to the surface energy 
from the marginally irrelevant
{\it bulk} operator actually gives rise to the leading singular 
temperature behavior
of the impurity susceptibility and the specific heat coefficient corresponding 
to a Curie-like behavior with logarithmic corrections.

We consider the antiferromagnetic spin-1/2 Heisenberg chain 
with open boundaries
\begin{equation}
H_{XXZ}=J\sum_{i=1}^N [S_{i}^xS_{i+1}^x+S_{i}^yS_{i+1}^y
+\Delta S_{i}^zS_{i+1}^z]
\label{xxz}
\end{equation}
in the limit $N \gg J/T$, i.e. the thermodynamic limit.
Note that in the opposite limit $T/J \ll 1/N$ the behavior is trivially
described by the ground state, which is a singlet for even $N$
with exponentially small susceptibility as $T\to 0$ and a doublet for odd $N$
with a Curie law behavior.  The crossover to the ground state
behavior is experimentally important and will be discussed later,  but for now
we will consider the low temperature behavior in the thermodynamic limit 
$1/N \ll T/J \ll 1$.
%We restrict our argument to the antiferromagnetic case $J>0$.
In the massless region, $0\leq\Delta \leq 1$, the low energy fixed point
of the Hamiltonian (\ref{xxz}) is the Tomonaga-Luttinger liquid, which belongs to
the universality class of the Gaussian theory with the central charge $c=1$.
The low energy effective Hamiltonian with leading irrelevant interactions
has been exactly obtained by Lukyanov\cite{luk}.
Using this effective theory
we can implement a perturbative expansion of the free energy
in terms of leading irrelevant interactions, 
and evaluate impurity corrections of order $1/N$. 
Following the idea of Cardy and Lewellen\cite{car1},
we consider a semi-infinite cylinder
on which the direction tangent to the circumference is taken as 
an imaginary time
axis, and the direction perpendicular to it as a space axis.
The circumference is equal to the inverse temperature $1/T$.
We define the boson phase field on this geometry, of which 
the mode expansion is given by
\begin{equation}
\phi^c(x,\tau)=Q-i\pi TPx+2\pi T\frac{w\tau}{\sqrt{2K}} %\nonumber \\
+\frac{i}{2}\sum_{n\neq 0}\frac{1}{n}
(\alpha_n e^{-2\pi Tn(x+i\tau)}+\bar{\alpha}_n
e^{-2\pi Tn(x-i\tau)}), \label{mode3}
\end{equation}
where $w$ is the winding number of $\phi^c$, 
$K$ is the Luttinger liquid parameter, and
the operators satisfy the commutation relations
$[\alpha_n,\alpha_m]=[\bar{\alpha}_n,\bar{\alpha}_m]=n\delta_{n+m,0}$,
$[\alpha_n,\bar{\alpha}_m]=0$, and $[Q,P]=i$.
Then, the low-energy effective Hamiltonian %for $1/2<\Delta<1$
on the semi-infinite cylinder is written as,
\begin{eqnarray}
H^c&=&H^c_0+H^c_{int}, \label{eff3}\\
H^c_0&=&
\int^{1/T}_0\frac{d\tau}{2\pi}[(\partial_{\tau}\phi^c)^2
+(\partial_x\phi^c)^2], \\
H^c_{int}&=&-a^{2K-2}\lambda\int^{1/T}_0\frac{d\tau}{2\pi}
\cos(\sqrt{8K}\phi^c).
\label{int1}
\end{eqnarray}
Here the constants $K$ and $a$
are parametrized as,
$K=[1-(1/\pi)\cos^{-1}(\Delta)]^{-1},$
$a=2(K-1)/[JK\sin(\pi/K)],$
%\begin{equation}
%\lambda=\frac{4\Gamma(K)}{\Gamma(1-K)}
%\left[\frac{\Gamma(1+1/(2K-2))}{2\sqrt{\pi}
%\Gamma(1+K/(2K-2))}\right]^{2K-2}.
%\end{equation}
and $\lambda$ is given by equation (2.24) in Ref.\onlinecite{luk}.
For this choice of the lattice parameter $a$, the velocity of spinons
is unity in the unit of $1/a$. Thus the Hamiltonian (\ref{eff3}) is obtained
by simply interchanging time and space coordinates of the usual
Hamiltonian defined on a circumference of $L=aN$. 
%The boson fields $\phi(x)$ and $\Pi(x)$ satisfy the canonical conjugate
%relation, $[\phi(x),\Pi(x')]=i\pi\delta(x-x')$.

%In the case of $0\leq \Delta \leq 1/2$ ($K\geq3/2$), 
%the low-temperature anomalous
%behaviors at boundaries do not appear, as is easily seen from
%the dimensional analysis.
%Thus we will not consider this case in the following.

%Generally, in systems with boundaries, there may be boundary operators
%in addition to bulk interactions.
%However, as was pointed out in ref.\cite{bru}, in the absence of
%symmetry-breaking external fields at boundaries, 
%we can exclude this possibility for the Heisenberg XXZ chains.

We express the partition function by using the transfer matrix 
$\exp(-LH^c)$
and the boundary state for $H^c_0$, $|B\rangle$.
The lowest order terms of the free energy for the Hamiltonian (\ref{eff3}) 
are given by,
%\end{multicols}
\begin{eqnarray}
F=-\frac{aT}{L}\ln\langle 0|e^{-LH^c_0}|B\rangle
+\frac{aT}{L}\int^L_0 dx\frac{\langle0|\exp(-LH^c_0)
\exp(xH^c_0)H^c_{int}\exp(-xH^c_0)|B\rangle}{\langle0|\exp(-LH^c_0)|B\rangle}
+...,
\label{free}
\end{eqnarray}
%\begin{multicols}{2}
\noindent
where $|0\rangle$ is the ground state of $H^c_0$.
The first term on the right-hand side of Eq.~(\ref{free}) is 
the free energy of the $c=1$ Gaussian model. 
The second term is the $1/L$ correction that
emerges as a result of boundary effects.
For the periodic boundary condition, this term vanishes.
In the case of an external magnetic field $h$, 
the free energy is evaluated by shifting the boson field $\phi^c(x)$
to $\tilde{\phi}^c(x)=\phi^c(x)-\sqrt{K/2}hx$.
Using Cardy and Lewellen's method\cite{car1}, we compute the first order
term as,
\begin{eqnarray}
-\frac{\lambda}{L}\frac{(\pi aT)^{2K}}{2\pi a}
\frac{\langle\Phi|B\rangle}{\langle 0|B\rangle}
\int^L_0dx\frac{\cos(2Khx)}{[\sinh(2\pi Tx)]^{2K}}, \label{bound}
\end{eqnarray}
where 
$|\Phi\rangle$ is the primary state that corresponds to
the conformal field $\exp(i\sqrt{8K}\phi^c)$. 
In the following, we evaluate Eq.~(\ref{bound}) exactly for
the free open boundary condition.

To calculate the prefactor in Eq.~(\ref{bound}),
we utilize properties of the boundary state.
A conformally invariant boundary condition is
imposed by demanding $T(z)=\bar{T}(\bar{z})$ at the boundary\cite{car2}.
Here $T(z)$ ($\bar{T}(\bar{z})$) is the holomorophic (anti-holomorophic) part
of the stress energy tensor.  
For the Gaussian model with $c=1$, this condition leads to
the following constraint on the boundary state\cite{ca1,ca2,osh}, 
\begin{eqnarray}
(\alpha_n\pm \bar{\alpha}_{-n})|B\rangle=0, \label{bc}
\end{eqnarray}
where the plus (minus) sign corresponds to the Neumann 
(Dirichlet) boundary condition.
Equation (\ref{bc}) is solved in terms of the 
Ishibashi states, which, in our case, are constructed from
the highest weight states of the U(1) Kac-Moody algebra $|v,w\rangle$
and their descendants\cite{ish}.
Here $v$, $w$ are integers specifying the U(1) highest weight state.
%The matrix element between $|0\rangle$ and $|B\rangle$ appeared in
%(\ref{bound}) can be computed by using these states.
The leading irrelevant interaction
(\ref{int1}) is expressed by 
the primary field $\exp(i\sqrt{8K}\phi^c)$ which 
corresponds to the primary state $|2,0\rangle$.
Therefore, $\langle \Phi |B\rangle$ is non-vanishing, only if $|B\rangle$
contains $|2,0\rangle$.  The Neumann boundary state which consists 
of the highest weight state $|0, w \rangle$ and its descendants
 does not satisfy this condition.
On the other hand, the Dirichlet boundary state,
\begin{eqnarray}
|D\rangle =\biggl(\frac{K}{2}\biggr)^{1/4}\sum_{v=-\infty}^{\infty}
e^{-i\sqrt{2 K}v\phi_0}
e^{-\sum_{n=1}^{\infty}\alpha_{-n}\bar{\alpha}_{-n}/n}
|v,0 \rangle ,  \label{dir}
\end{eqnarray}
has a finite overlap with $| \Phi \rangle$.
%In the following, we put $\phi_0=0$.
%This choice of the averaged value of the phase at the boundary
%is consistent with the free open boundary condition for 
%the boundary magnetization.
%This choice of the averaged value of the phase at the boundary
%does not affect boundary quantities.
Then, we have,
\begin{eqnarray}
\frac{\langle 2,0 |D\rangle}{\langle 0| D\rangle}=1. \label{prefac}
\end{eqnarray}
Carrying out the integral in Eq.~(\ref{bound}) and using Eq.~(\ref{prefac}), 
we obtain corrections to the boundary part of the free energy,
\begin{eqnarray}
\delta F_{\rm B}=-\frac{\lambda}{2\pi aN}(2\pi a T)^{2K-1}{\rm Re}
[B(K+i\frac{Kh}{2\pi T},1-2K)], \label{freeb}
\end{eqnarray}
where $B(x,y)=\Gamma(x)\Gamma(y)/\Gamma(x+y)$.

The boundary contribution to the spin susceptibility is derived from 
Eq.~(\ref{freeb}),
\begin{eqnarray}
\chi_{\rm B}=\frac{\lambda aK^2}{2\pi N}
B(K,1-2K)[\pi^2-2\psi'(K)](2\pi a T)^{2K-3}, \label{chi1}
\end{eqnarray}
with $\psi'(x)=d\psi(x)/dx$. 
Note that for $1<K<3/2$ ($1/2<\Delta<1$), the boundary spin
susceptibility $\chi_{\rm B}$ shows a divergent behavior $\sim 1/T^{3-2K}$,
as temperature decreases.
This anomalous temperature dependence is also observed in the boundary part 
of the specific heat coefficient computed as,
\begin{eqnarray}
\frac{C_{\rm B}}{T}
=\frac{2\pi a\lambda}{N}(2K-1)(2K-2)B(K,1-2K)(2\pi a T)^{2K-3}.
\label{heat1}
\end{eqnarray}
Boundary terms that are regular in $h$ and $T$ give higher order 
corrections and have been neglected here.
We would like to stress that in the formulas (\ref{chi1}) and (\ref{heat1})
there is no free parameter, and the prefactors are exact.
The divergent behaviors  of Eqs.~(\ref{chi1}) and (\ref{heat1})
for $T\rightarrow 0$ are physically understood as follows.
In contrast to the bulk Heisenberg chains, 
the ground state degeneracy at the boundaries gives rise to 
large spin fluctuations, which disturb the spin singlet formation.
It should be emphasized that
the singular behaviors are not due to the presence of boundary operators,
but interpreted as a consequence of finite-temperature corrections of
the surface energy and the boundary entropy $\ln\langle 0|B\rangle$ 
caused by {\it bulk} irrelevant interactions.

At zero temperature with a small magnetic field,
a similar singular behavior appears in the field dependence of
the boundary spin susceptibility given by,
\begin{eqnarray}
\chi_{\rm B}(T=0)=\frac{\lambda(aK)^{2K-1}}{2\pi aN}\sin(\pi K)
\Gamma(3-2K)h^{2K-3}. \label{zerospin}
\end{eqnarray}
The zero temperature susceptibility is also derived from
the Bethe ansatz exact solution by using the Wiener-Hopf method.
We have checked that Eq.~(\ref{zerospin}) coincides with the result obtained
by the Bethe ansatz method. 

Now let us consider the isotropic case $K=1$.
The free energy correction (\ref{freeb}) possesses poles for $K=1$.
To deal with these singularities, we follow the procedure
considered by Lukyanov for bulk spin systems\cite{luk}.
We rewrite $H_{int}$ in terms of the SU(2) current operators,
\begin{eqnarray}
H_{int}=\int\frac{dx}{2\pi}[g_{\parallel}J_0\bar{J}_0
+\frac{g_{\perp}}{2}(J_{+}\bar{J}_{-}+J_{-}\bar{J}_{+})].
\end{eqnarray}
The exact expressions for the running coupling constants 
are known as $g_{\parallel}=2(1-1/K)(1+q)/(1-q)$, 
$g_{\perp}=4(1-1/K)q^{1/2}/(1-q)$\cite{luk,za}.
Here the parameter $q$ is the function of $T$ and $h$, of which the expression
is also known exactly.
On the other hand, for the value of $K$ close to 1, 
Eq.~(\ref{freeb}) can be expanded in power series of $1-1/K$.
Comparing the expansion of Eq.~(\ref{freeb}) 
with the expression for $g_{\parallel}$ and $g_{\perp}$,
we can write the free energy correction (\ref{freeb}) as a power series 
expansion in terms of $g_{\parallel}$ and $g_{\perp}$.
Then, taking the limit $K\rightarrow 1$ and 
$g_{\parallel},g_{\perp}\rightarrow g$,
we end up with,
\begin{eqnarray}
\delta F_{\rm B}=-\frac{Tg}{2N}-\frac{h^2}{24NT}(g+g^2)+.... \label{free2}
\end{eqnarray}
for $h\ll T$.
The running coupling constant is determined from the equation,
\begin{eqnarray}
g^{-1}+\frac{1}{2}\ln(g)=-{\rm Re}[\psi(1+\frac{i h}{2\pi T})]
+\ln(\sqrt{\frac{\pi}{2}} \frac{e^{1/4}J}{T}), \label{sca}
\end{eqnarray}
where $\psi(x)$ is the di-gamma function.
Using Eqs.~(\ref{free2}) and (\ref{sca}), we obtain 
the leading term of the boundary spin susceptibility and
the specific heat coefficient,
\begin{eqnarray}
\chi_{\rm B}&=&\frac{g}{12NT}+\frac{g^2}{12NT}(1-\frac{3\psi''(1)}{2\pi^2})
+... %\nonumber \\
= \frac{1}{12NT\ln(\alpha/T)}\biggl(1-\frac{\ln\ln(\alpha/T)}{2\ln(\alpha/T)}
+...\biggr), \label{chi2}
\end{eqnarray}
\begin{eqnarray}
\frac{C_{\rm B}}{T}=\frac{g^2}{2NT}+\frac{5g^3}{4NT}+...= 
\frac{1}{2NT(\ln(\alpha/T))^2}\biggl(1-\frac{\ln\ln(\alpha/T)}{\ln(\alpha/T)}
+...\biggr),
\label{spe2}
\end{eqnarray}
where %$\psi''(1)=-2.40411...$, and 
$\alpha=\sqrt{\pi/2}\exp(1/4+\gamma)$
with $\gamma$ the Euler constant.
At zero temperature, the boundary spin susceptibility 
for a small magnetic field
is also derived from the opposite limit $h \gg T$ of Eq.~(\ref{freeb}),
\begin{eqnarray}
\chi_{\rm B}(T=0)=\frac{g^2}{4Nh}+\frac{5g^3}{8Nh}+...
=\frac{1}{4Nh(\ln(2\pi \alpha/h))^2}
\biggl(1-\frac{\ln\ln(2\pi\alpha/h)}{\ln(2\pi\alpha/h)}
+...\biggr).\label{zero}
\end{eqnarray}
This result (\ref{zero}) coincides completely
with that obtained by the Bethe ansatz exact solutions\cite{ess,as,fra,fuji2}.

The prefactor of the leading terms  in
 Eqs.~(\ref{chi2}), (\ref{spe2}), and (\ref{zero}) are exact for the 
isotropic Heisenberg chain, but not universal.   For example a frustrating
nearest neighbor coupling $J_2$ will change the prefactors and at the critical 
point $J_2 \approx 0.241167$\cite{egg3} the singular behavior is absent.
Special logarithmic singularities at boundaries 
have also been found in the NMR relaxation rate\cite{qin,bru},
which are also directly related to the bulk marginal operator.
However, in Eqs.~(\ref{chi2}) and (\ref{spe2}) it is even the 
leading T-dependence that is changed due to this irrelevant operator.

%A logarithmic singularity similar to (\ref{chi2}) and (\ref{spe2})
%was also found before in the NMR relaxation rate
%at an edge of spin chains calculated by Affleck and Qin, and Brunel et al., 
%who showed that the bulk marginal interaction
%gives rise to an anomalous dimension at boundaries.\cite{qin,bru}
%It is noted that the logarithmic singularity found in this paper
%has the same origin as that for
%the NMR relaxation rate at boundaries.

\begin{figure}[h]
\centerline{\epsfxsize=13.6cm \epsfbox{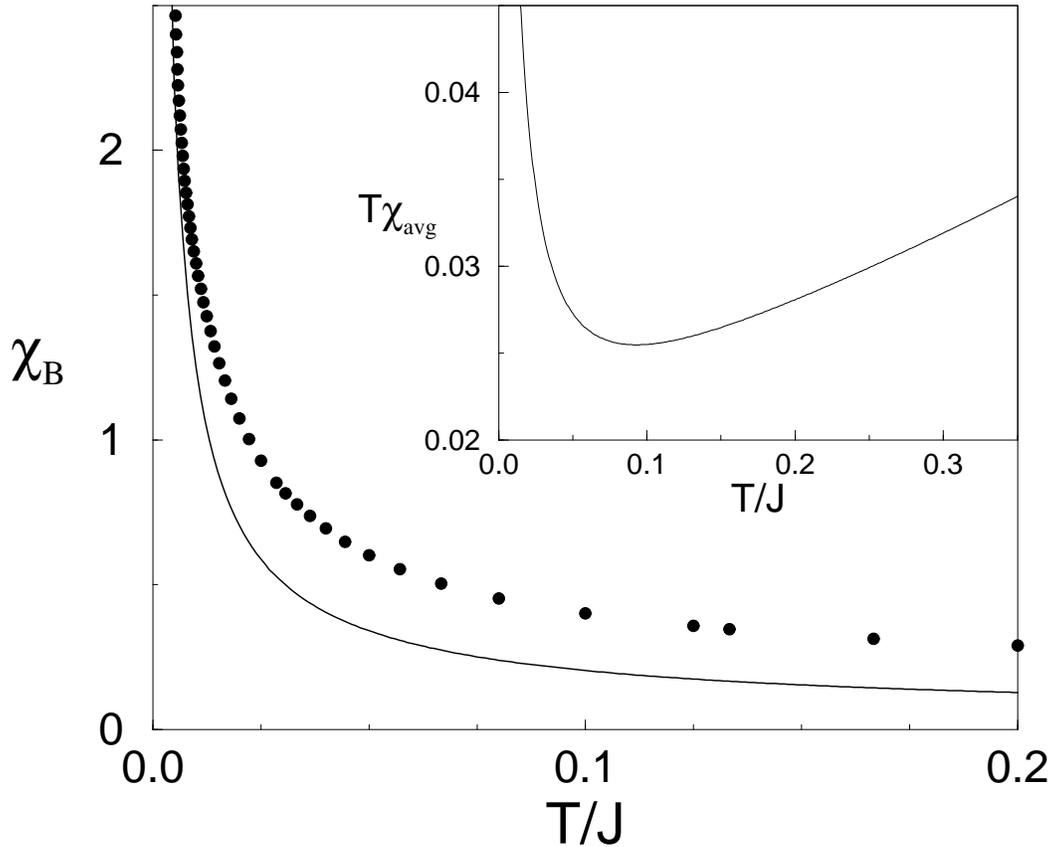}}
\caption{The boundary susceptibility $\chi_{\rm B}$ as a function of $T$
according to the DMRG calculations (black points),
compared to the result of the field theory in Eq.~(\ref{chi2}).
Inset: approximate temperature dependence of the average Curie constant per impurity for
an impurity density of $\rho=0.1\%$} \label{chifig}
\end{figure}

For comparison we have also used the numerical density matrix renormalization 
group for transfer matrices (TMRG) applied to impurity problems\cite{rom}.  
This method can calculate local expectation values and the impurity free 
energy directly in the thermodynamic limit $N\to \infty$.  When evaluating 
the impurity susceptibility we need to take the second derivative of the 
impurity free energy and therefore subtract two large numbers, which becomes 
inaccurate for very low temperatures.  For the lowest temperatures ($T<0.1J$)
we have therefore instead summed over the excess local responses in a range 
around the open ends, which gave more accurate results. 
This method agrees with 
taking the second derivative for higher $T$ and should also be a good 
approximation for $T<0.1J$. Note, however, that 
taking the excess response only at the site closest to the open end as
was done in Ref.~\onlinecite{rom} gives a weaker temperature dependence.
The results are shown in Fig.~\ref{chifig} without any adjustable parameters.

%The results are shown in Fig.~\ref{chifig} with the plot of
%the field theoretical result (\ref{chi2}).
%We see that they coincide well with each other.

Finally, we would like to remark on the implications of our findings for
experimental observations.   In experimental quasi-one dimensional 
systems such as ${\rm Sr_2CuO_3}$ a small density $\rho$
of impurities is always present in 
the form of intrinsic defects of the crystal which effectively cut 
the chains. A corresponding Curie contribution has been observed that
can be strongly reduced by careful annealing\cite{uchi,ami},
which implies that this Curie tail cannot be due to magnetic impurities
in the sample.  
For extremely low temperatures $T/J \ll 1/N$ such a Curie behavior can be explained by 
finite chains with odd $N$ that have locked into their doublet ground state\cite{haas}.
We have now shown that a Curie-like behavior can even be expected
for higher temperatures, albeit with a different Curie constant 
in Eq.~(\ref{chi2}).
The crossover between the ground state and the thermodynamic behavior
is well understood, and in fact the partition function can be written
explicitly if irrelevant operators are ignored\cite{egg}.  Roughly,
 the crossover 
occurs when the temperature becomes comparable to the finite size
energy gap $T \sim \pi v/N$.  For a carefully annealed sample of 
${\rm Sr_2CuO_3}$ with $\rho \sim 0.013\%$ \cite{uchi} this means
that the ground state contribution is only significant 
for $T \alt 0.001J \sim 2K$. For higher temperatures 
the impurity susceptibility is dominated by the expression in Eq.~(\ref{chi2}).
Experimentally this means that the effective Curie constant first drops logarithmically as
the temperature is lowered. 
In previous studies this slight change of the Curie constant has been 
fitted to a Curie Weiss correction\cite{ami} assuming a 
phenomenological interaction between the 
impurity moments. The result in Eq.~(\ref{chi2}) now gives a clear 
prescription for the expected form of the impurity susceptibility.
At the lowest temperatures $T \alt 5\rho J$ the Curie constant increases 
again sharply to
a limiting value of $\rho/8J$ as $T\to 0$ due to the ground state 
contributions of the chains with odd $N$\cite{haas}, leading to a characteristic 
minimum in the average Curie constant $T\chi_{\rm avg}$.  
The approximate behavior of the averaged impurity 
susceptibility $T \chi_{\rm avg}$ is shown in 
the inset of Fig.~\ref{chifig} for $\rho = 0.1\%$ by averaging over all 
chain lengths
and assuming a sharp crossover from ground state to thermodynamic behavior.

In summary, we have studied the boundary thermodynamics for
spin-1/2 Heisenberg chains with open ends
by using boundary conformal field theory and numerical TMRG calculations.
It has been shown that the boundary contributions of the
spin susceptibility and the specific heat coefficient 
exhibit divergent behaviors as the temperature is lowered, 
which explains experimental observations in quasi one dimensional compounds.

We would like to thank I.~Affleck, F.H.L.~Essler, and 
N.~Kawakami for valuable discussions.
We are also grateful to A.~Furusaki for sharing his insight about
related issues which led us to find an error in an earlier version of 
the formula (18).
This work was partly supported by a Grant-in-Aid from the Ministry
of Education, Science, Sports and Culture, 
Japan and the Swedish Research Council.

\end{document}